\documentclass[10pt]{iopart}

\usepackage[numbers,sort&compress]{natbib}
\usepackage{lipsum}
\usepackage{graphicx}
\usepackage{hyperref}
\usepackage{siunitx}

\begin{document}

\title[Low-loss mm-wave resonators with improved coupling]{Low-loss millimeter-wave resonators with an improved coupling structure}

\author{
A Anferov$^{12}$,
S P Harvey$^{34}$,
F Wan$^{34}$,
K H Lee$^{12}$, 
J Simon$^{3}$ and 
D I Schuster$^{34}$}

\address{$^1$ James Franck Institute, University of Chicago, Chicago, IL 60637, USA}
\address{$^2$ Department of Physics, University of Chicago, Chicago, IL 60637, USA}
\address{$^3$ Department of Applied Physics, Stanford University, Stanford, CA 94305, USA}
\address{$^4$ SLAC National Accelerator Laboratory, Menlo Park, CA, 94025 USA}
\eads{\mailto{dschus@stanford.edu}, \mailto{aanferov@uchicago.edu}}
\begin{indented}
\medskip
\item[]\today 
\end{indented}
\begin{abstract}
Millimeter-wave superconducting resonators are a useful tool for studying quantum device coherence in a new frequency domain.
However, improving resonators is difficult without a robust and reliable method for coupling millimeter-wave signals to 2D structures. 
We develop and characterize a tapered transition structure coupling a rectangular waveguide to a planar slotline waveguide with better than 0.5 dB efficiency over 14 GHz, and use it to measure ground-shielded resonators in the W band (75 - $110~$GHz).
Having decoupled the resonators from radiative losses, we consistently achieve single-photon quality factors above $10^5$, with a two-level-system loss limit above $10^6$, and verify the effectiveness of oxide removal treatments to reduce loss.
These values are 4-5 times higher than those previously reported in the W band, and much closer to typical planar microwave resonators.
The improved losses demonstrated by these on-chip millimeter-wave devices shed new light on quantum decoherence in a different frequency regime, offer increased selectivity for high-frequency detectors, and enables new possibilities for hybrid quantum experiments integrating millimeter-wave frequencies.
\end{abstract}

\maketitle
\ioptwocol

\section{Introduction}
Extending superconducting quantum device functionality to millimeter-wave frequencies (near $100~$GHz) offers new opportunities for detection and transduction \cite{pechalSafavi-Naeini2017} as well as access to large coupling strengths for hybrid quantum experiments \cite{xiangNori2013,clerkNakamura2020}. 
Most importantly, the reduced sensitivity to thermal noise of higher-energy millimeter-wave photons could enable quantum experiments at 1~K, which significantly reduces cooling complexity and power dissipation constraints \cite{Pobell2007}, enabling new pathways for scaling up quantum computing platforms, and could facilitate direct integration with high-speed superconducting digital logic \cite{leonard2019sfqControl,mcdermott2014sfqControl}.
High-frequency superconducting devices have already found uses in sensitive detectors \cite{tucker1985millimeterrev,banys2022mmKIPA,levinsen1980mmJPA} with applications in high-frequency radio astronomy and potentially molecular spectroscopy \cite{aslam2015mmesr,ishikawa2018mmesr,vasilyev2004mmesr}.
In particular, superconducting resonators are well suited for millimeter-wave filter applications \cite{gaoLeduc2009,endoKlapwijk2013,shirokoffZmuidzinas2012} making narrower line-width devices desirable.

To improve detector selectivity and sensitivity, and to establish more robust high-frequency quantum information systems, it is vital to understand and minimize decoherence in superconducting devices.
Significant effort has gone towards investigating and reducing sources of loss at microwave frequencies \cite{mcdermott2009}, establishing that a significant remaining contribution to decoherence at the single photon level comes from two-level-systems (TLS) found in amorphous dielectric materials.
At temperatures near 1~K, these limiting loss factors can be overshadowed by dissipation from quasiparticles \cite{mattisBardeen1958,catelani2011quasiparticle} in commonly used superconductors such as aluminum or tantalum. 
Using a higher critical temperature ($T_c$) superconductor such as niobium avoids generating quasiparticles with millimeter-wave photons and mitigates loss from thermal quasiparticles.
However, since probing individual decoherence mechanisms requires reducing all other sources of loss as well, the nature and limits of TLS loss contributions and their frequency dependence when scaled to millimeter-wave frequencies remain relatively unexplored.

When measured at single-photon energies, the internal quality factor ($Q_i$) of an on-chip resonant circuit provides insight into the maximum coherence of a quantum system formed by adding a source of nonlinearity (such as kinetic inductance \cite{stokowskiSafavi-Naeini2019,Faramarzi2021kineticon,anferovSchuster2020} or a high-frequency Josephson junction \cite{kimSemba2021,anferovSchuster2023b}).
Significant progress has been made in improving millimeter-wave resonators \cite{stokowskiSafavi-Naeini2019,anferovSchuster2020,gaoLeduc2009,shirokoffZmuidzinas2012,endoKlapwijk2013,u-yenWollack2017}, but their single-photon quality factors remain below 2-$4\times10^4$: significantly lower than those measured in microwave circuits \cite{mcraeMutus2020}.
This can largely be attributed to two primary factors: first, millimeter-wave resonators frequently use materials with low dielectric constants (such as SiO$_2$) to simplify high-frequency circuit design \cite{gaoLeduc2009,jia2019mmfilter,Kerr2014alma}, despite their poor dielectric loss characteristics \cite{Krupka1999losstan} compared to crystalline silicon or sapphire.
Second, many existing millimeter-wave resonators rely on coplanar stripline or microstrip transmission line components, both of which have relatively high radiation profiles resulting in increased radiative losses, particularly at high frequencies \cite{vanberkel2015}.

Taking inspiration from low-loss microwave devices, a ground-shielded circuit design minimizing radiative loss offers an attractive method for increasing millimeter-wave quality factors, and better control over on-chip signal propagation and coupling. 
Whereas low-frequency signals can be routed to an on-chip waveguide directly through wire bonds, these introduce substantial parasitic inductance \cite{Wenner2011wirebond}. Grounded circuits also present additional design challenges in millimeter-wave bands, where signals are primarily transmitted by hollow waveguides, requiring a method to efficiently direct the waveguide electromagnetic fields onto the chip.
This proves to be a challenging problem, and as a result, a variety of transition structures coupling waveguides with on-chip transmission lines have been developed \cite{jastrzebskiLi2002}; however, with no universal solution, transitions for specific applications are still actively studied and improved. 

In this work, we use a transition specifically designed to measure ground-shielded millimeter-wave resonant circuits with improved control.
We characterize a tapered coupling structure that efficiently confines the signal fields to an on-chip slotline waveguide, finding an insertion loss better than 0.5 dB over 14 GHz of bandwidth.
We study niobium resonators patterned near the slotline and show that resonator coupling can be controlled independently without increasing radiation loss.
With this novel resonator design combined with fabrication procedure, we improve on existing planar millimeter-wave devices and achieve high internal quality factors ($Q_i$) consistently exceeding $10^5$ at single photon powers.
This allows us to study the effects of oxide growth and removal on remaining loss contributions to $Q_i$ from millimeter-wave two-level systems (TLS), and show that limits from TLS can be increased as high as $10^6$.

\begin{figure}[htb]
\centering
\includegraphics[width=3.28in]{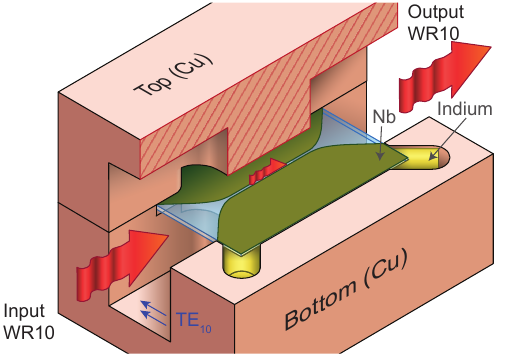}
\caption{Cutaway diagram showing assembled back-to-back waveguide to on-chip slotline transition structures, with signal propagation marked in red. 
In the slotline region the waveguide geometry is constricted to prevent spurious propagating modes.
On the chip corners, rounded channels allow indium wire (yellow) to deform, which secures the chip in place.
}
\label{fig1}
\end{figure}

\section{Tapered Waveguide Transition Design}

When designing a millimeter-wave circuit to minimize loss, additional constraints apply to its coupling structure.
Crystalline sapphire is an ideal substrate choice as it yields much lower dissipation than most dielectrics, but is more difficult to machine which imposes design limitations. 
As a result, transition designs requiring abnormally cut non-rectangular substrates \cite{vallettiDiPaolo2022,yassinNorth2008, jingZhengYan-qiu2013}, micromachining \cite{leeKatehi2004} or drilled holes \cite{alhenawySchneider2012} are impractical.
Sapphire also presents additional challenges for high frequency circuit design due to its relatively high dielectric constant, which leads to more pronounced impedance mismatches caused by the presence of substrate in the waveguide \cite{yipLi2002}.
Ensuring currents are carried by superconducting materials minimizes conduction loss.
To achieve this, on-chip superconducting layers should be well separated from both waveguide and housing metal \cite{huangOliver2021}: this consideration makes transitions with stripline geometries, which needs an external ground plane \cite{wuXu2022}, less ideal.
While potentially offering low radiation loss, coplanar waveguide transitions \cite{mottonenRaisanen2004,leeKatehi2004} are typically more complex, requiring multiple stages and more physical space.
Finline transitions, on the other hand, consist of a single taper \cite{gieseJacob2015,yassinNorth2008,jingZhengYan-qiu2013}, making them more compact, and transform the signal into a differential mode localized on the chip surface.

\begin{figure}[htb]
\centering
\includegraphics[width=3.28in]{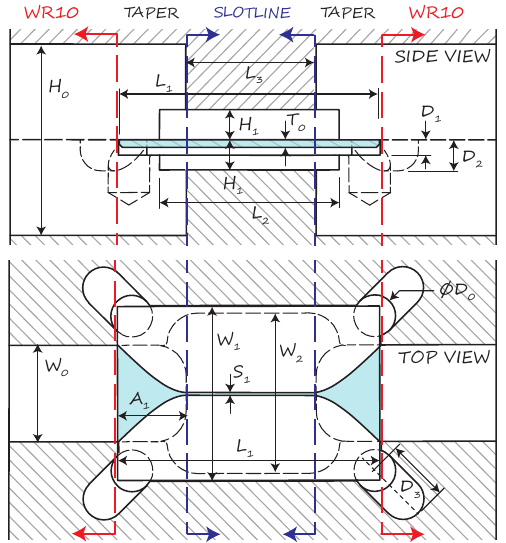}
\caption{Side and top section views of the transition structure geometry, with reference planes and relevant dimensions marked. Un-metallized sections of the chip are shown in blue. The rounded corner channels are completely filled by indium.}
\label{fig2}
\end{figure}
\begin{table}[htb]
\caption{\label{tab1}Optimized taper dimensions used (in mm).}
\begin{indented}
\item[]\begin{tabular}{@{}lllllllll}
\br
$H_0$&$L_1$&$L_2$&$L_3$&$T_0$&$H_1$&$D_1$&$D_2$\\
\mr
2.54&3.45&2.27&1.66&0.1&0.4&0.2&0.412\\
\mr
$W_0$&$W_1$&$W_2$&$S_1$&$A_1$&$D_0$&$D_3$&$R_1$\\
\mr
1.27&2.29&2.11&0.04&0.895&0.55&0.8&0.4\\
\br
\end{tabular}
\end{indented}
\end{table}

Our transition is defined by a superconducting niobium film patterned on the top surface of a rectangular crystalline sapphire substrate centered in a rectangular waveguide.
Inspired by Refs. \cite{gieseJacob2015,yassinNorth2008,alhenawySchneider2012}, the geometry consists of a differential unilateral finline that tapers from the waveguide width down to a narrow slotline.
A cutaway diagram of the complete structure is shown in \Fref{fig1}, with two transitions coupled back-to-back by a length of on-chip slotline useful for coupling to resonators.
Where the transition terminates, the waveguide height is reduced to increase the cutoff frequency of unwanted higher-order propagating modes, while the waveguide width is broadened to increase the usable area of the chip: this allows the device area to remain suspended away from metal surfaces, which minimizes conduction loss \cite{huangOliver2021}.
The only direct contact with the copper enclosure occurs where the chip is clamped at its corners. 

\begin{figure*}[ht]
\centering
\includegraphics[width=6.7in]{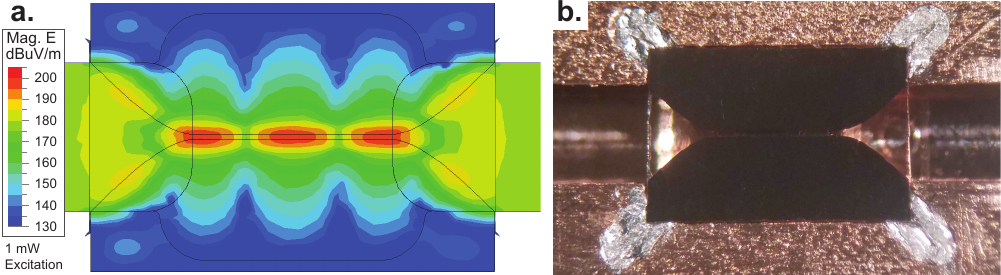}
\caption{\textbf{a)} Simulated electric field distribution (log scale) viewed from the top for a wave traveling through the structure. Notably, a much greater dynamic range of electric field strengths is achieved as the signal is compressed into the slotline, as compared to just the waveguide field.
\textbf{b)} Photograph of a mounted chip with back to back transition structures, with top copper block removed. The indium, visible on the corners of the chip, deforms and fully fills the designated channels, thermalizing and securing the chip.}
\label{fig34}
\end{figure*}

Due to the pronounced impedance discontinuity between the sapphire chip and the waveguide \cite{yipLi2002}, a broadband matching structure is difficult to achieve with an exponential or cosine Vivaldi taper contour as is typically used in finline transitions \cite{gieseJacob2015,alhenawySchneider2012}.
Instead, we find that a curved taper shape with a nearly linear waveguide entrance can compensate for the mismatch and can be optimized to give good performance over a section of waveguide bandwidth.
The optimized contour is described by the function:
\begin{equation}
y(x) = \left(\frac{W_0-S_1}{2}\right)\frac{x}{A_1}\sqrt{2-\left(\frac{x}{A_1}\right)^2}
\end{equation}
where $W_0$ is the smaller waveguide dimension, $S_1$ is the slotline width, and $A_1$ is the transition length.
The geometry of the transition is detailed in \Fref{fig2}.
Using finite element method simulation software\footnote{Ansys HFSS}, the above contour function and parameterized geometry dimensions are optimized for maximal transmission in the 90-$100~$GHz band.
The resulting optimal dimensions are listed in \Tref{tab1}.
Notably, by relaxing the bandwidth optimization constraint, we achieve a taper structure less than $0.9~$mm long: much smaller than the $\lambda/2$ value predicted with analytical functions \cite{yipLi2002} and more compact than many other transition structures in literature \cite{alhenawySchneider2012,yassinNorth2008,jingZhengYan-qiu2013,hungWu2005,gieseJacob2015}.

The benefit of coupling to resonant devices through an intermediate on-chip slotline transmission line is apparent when examining the magnitude of the electric field of a wave propagating through the structure, shown in \Fref{fig34}a.
Whereas some previous implementations of millimeter-wave resonators \cite{anferovSchuster2020,stokowskiSafavi-Naeini2019} interact with a uniform waveguide electric field (visible on the left and right ends of \Fref{fig34}a) and rely on varying resonator dipole moments to adjust coupling, our method compresses the signal to a $40~\mu$m slotline resulting in over $50~$dB of dynamic range in electric field strength across the usable area of the chip.
Consequently, the dipole coupling strength for each resonator can be set by its location on the chip, without needing to adjust the resonator's dipole moment, leaving more freedom to optimize the resonator performance.

\section{Waveguide Transition Characterization}
The taper geometry described above is defined by chlorine reactive ion etching a $100~$nm thick film of high-purity electron-beam-deposited niobium grown on a $100~\mu$m thick crystalline C-plane sapphire substrate (see Appendix A for detailed procedure).
Short sections of superconducting indium wire which form robust cryogenic seals compatible with low-loss superconducting devices \cite{brecht2017micromachined}, are used to secure and thermalize the chip to the copper waveguide enclosure, visible in a photograph of the mounted chip shown in \Fref{fig34}b.
The edges of the on-chip taper structure match the waveguide dimensions, which enables visual alignment during mounting (necessary to maximize coupling). 

\begin{figure}[b]
\centering
\includegraphics[width=3.28in]{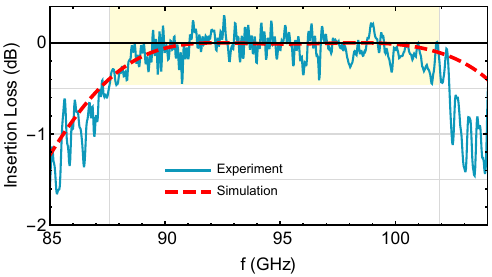}
\caption{De-embedded insertion loss for two back-to-back transitions along with simulated values. In the band of interest (highlighted), we find an insertion loss better than 0.46 dB, limited by de-embedding calibration uncertainty ($\sim0.3~$dB).}
\label{fig5}
\end{figure}
\begin{figure*}[ht!]
\centering
\includegraphics[width=5.0in]{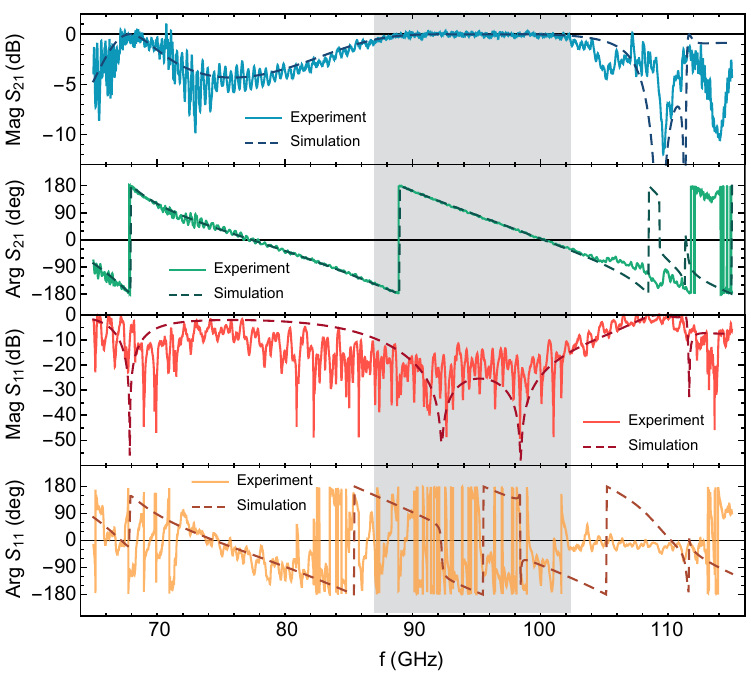}
\caption{Wideband de-embedded scattering matrix parameter measurements (solid lines) for two transitions back-to-back along with respective simulations (dashed lines) showing good agreement. In the operating band marked in gray, we find a total insertion loss better than 0.5 dB, and return loss of $13.1\pm8.6~$dB.}
\label{fig6}
\end{figure*}

The assembled structure is then measured at $0.86~$K in a helium-4 cryostat: using a vector network analyzer with millimeter-wave extension modules and a cryogenic low noise amplifier, we measure the complex response in transmission and reflection.
Input attenuation and cryogenic isolators reduce thermal noise reaching the sample, enabling measurements in the single photon limit.
A manual calibration removes the effects of additional hardware on the input and output lines, recovering de-embedded scattering parameters of the sample.
Cryogenic calibrations are a complex problem \cite{Yeh2013cryocal,Wang2021cryocal,Cataldo2015cryocal} however due to mechanical shifts from thermal cycling and the large amounts of attenuation and gain present in the measurement network (see Appendix B). This yields higher levels of uncertainty in our measured scattering parameters.

The effectiveness of the taper transitions is tested by using transitions to convert a waveguide signal to a slotline and back.
The measurement results are summarized in \Fref{fig5} and \Fref{fig6}.
We find the transition performs best between $87.6-102.4~$GHz, exceeding the designed range, and define this $14.8~$GHz wide frequency range as the useful operating band.
Within the operating band, we find a maximum insertion loss of approximately $0.46\pm{0.35}~$dB (or 94.8\% transmission), corresponding to a coupling efficiency of $\sim0.23~$dB for a single taper structure.
However these values are likely dominated by errors introduced by calibration methods (which do not enforce passivity).

Across the W band, we find that the de-embedded transmission and reflection of the structure are in fairly good agreement with simulations, with the exception of the region near $110~$GHz: in this region unwanted resonances occur in the substrate and indium mounting regions, which are difficult to predict.
For better performance in this range, our geometry could be adjusted to shift these resonances even higher in frequency, or reduce the energy participation in the indium regions.
In the operating band, we find the return loss exceeds $13.1\pm8.6~$dB, which slightly deviates from simulation, but could be attributed to the significantly increased calibration uncertainty due to the high transparency of the structure and increased sensitivity to error terms.
Combined, these measurements demonstrate a transition structure in good agreement with simulation, which in the operating band couples a signal on and off a chip with high efficiency.

\section{Ground-Shielded Resonator Design}
\begin{figure}[ht]
\centering
\includegraphics[width=3.1in]{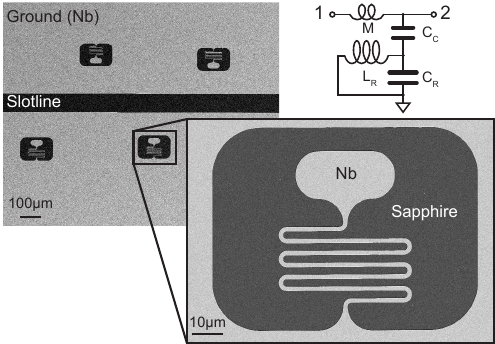}
\caption{Electron micrograph of resonator geometry and coupling arrangement relative to feedline. This structure can be approximated by a simplified circuit diagram (top-right).}
\label{fig7}
\end{figure}
\begin{figure}[hb!]
\centering
\includegraphics[width=2.8in]{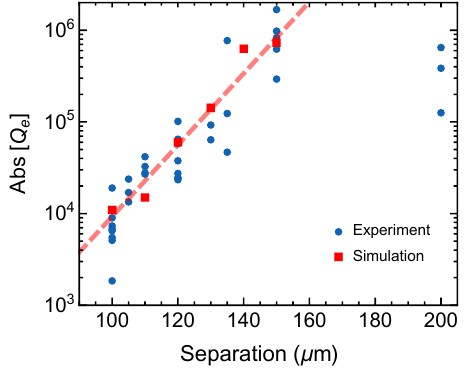}
\caption{Simulated (red) resonator coupling $Q_e$ as a function of separation from the slotline with an empirical fit (red line) used for predictions. Experimental measurements of $Q_e$ (blue) are in reasonable agreement.}
\label{fig8}
\end{figure}

Having demonstrated a coupling structure capable of efficiently transforming a rectangular waveguide signal to and from a localized on-chip slotline, we can now design resonators decoupled from their environment while only considering local interactions.
Away from the centered slotline, the chip surface is entirely covered by superconducting material acting as a ground plane, so a resonant structure patterned in this region will have its long-range electric dipole interactions reduced.
Our millimeter-wave resonator design is composed of a discrete capacitor island connected to the ground plane by a $2~\mu$m-wide meandered inductor, shown in \Fref{fig7}.
Every resonator is designed with an identical capacitor island, and its resonant frequency $\omega_0$ is adjusted by changing the inductor length while keeping its width constant.
The entire resonator has a rectangular footprint around $100-200~\mu$m per side, which is not insignificant compared to signal wavelength in the slotline ($\sim1~$mm): in this limit, the ground plane edges contribute significant reactive corrections.
On each chip, five to six resonators are placed near the central slotline to allow interaction with the propagating signal.
This differential geometry can be modelled by the single-ended $LC$ circuit shown in \Fref{fig7} including asymmetric coupling ($C_C$ and $M$) \cite{khalilOsborn2012}. This accounts for reactive contributions and impedance mismatches induced by the resonator presence.

For this resonator geometry, coupling strength can be controlled by adjusting its separation from the slotline.
The interaction decreases exponentially with distance, which we demonstrate by simulating the coupling quality factor $Q_e$ as a function of resonator separation, plotted in red in \Fref{fig8}.
The coupling quality factor can also be  approximated empirically as a function of only the separation $d$ by  $\log_{10} Q_e=0.06491+0.0390 d/\mu$m, shown as a dashed line in \Fref{fig8}.
We find that experimental measurements of $Q_e$ (described in the next section) follow this approximation reasonably well.

\begin{figure*}[ht]
\centering
\includegraphics[width=6.7in]{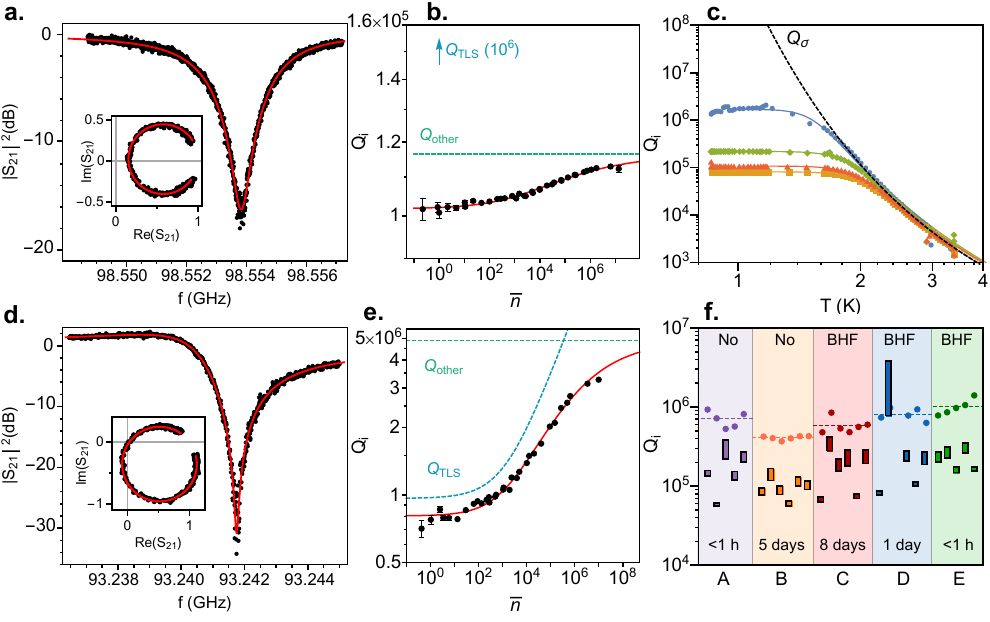}
\caption{\textbf{a-b)} Complex transmission spectrum of a typical resonator, and power dependence of its internal quality factor along with fits to a model including TLS and independent loss (red). Here, $Q_i$ is primarily limited by non-TLS loss ($Q_\text{other}$).
\textbf{c)} Temperature dependence of $Q_i$ at $\bar{n}\sim10^5$ for resonators from chip D.
The black dashed line corresponds to a Bardeen-Cooper-Schrieffer (BCS) model of conductivity loss, and solid lines are respective fits to a model including conductivity and TLS loss. 
\textbf{d-e)} Complex transmission spectrum and $Q_i$ power dependence of the best resonator measured in this study.
For this device, $Q_\text{TLS}$ is the dominant loss source.
\textbf{f)} Internal quality factors for resonators in this study, grouped by etching conditions and elapsed time after fabrication.
The top and bottom of the colored bars correspond to measured low-power and high-power limits of $Q_i$, and the points correspond to TLS induced loss $Q_{\mathrm{TLS},0}$ with averages for each chip denoted by a dashed line.}
\label{fig9}
\end{figure*}
\section{Millimeter-wave Resonator Measurements}
Characterizing the complex transmission spectra of these resonators at low temperatures ($0.86~$ K) allows us to explore losses at millimeter-wave frequencies.
Typical normalized measurements at low average photon number ($\bar{n}\approx10$) are shown in \Fref{fig9}a.
On resonance, we observe a dip in transmission as the resonance sweeps through a circle in the complex plane. This behavior is captured well by \cite{khalilOsborn2012}:
\begin{equation}
    S_{21}=1-\frac{Q}{Q_e^*}\frac{e^{i\phi}}{1 + 2i Q \frac{\omega-\omega_0}{\omega_0}}
\label{eq_s21}
\end{equation}
where $Q^{-1}=Q_i^{-1}+\text{Re}[Q_e^{-1}]$ \cite{khalilOsborn2012} and the coupling quality factor $Q_e$ is rotated in the complex plane by $Q_e= Q_e^* e^{-i\phi}$ due to asymmetric coupling to the slotline as described in the previous section.
Measuring both quadratures of the transmission spectrum to capture this asymmetry is particularly important for extracting an accurate estimate of $Q_i$, which is sensitive to $\phi$ \cite{probstWeides2015}.

Repeating these measurements at varying powers shows that $Q_i$ increases with power, as shown in \Fref{fig9}b.
This increase can be explained by a power-dependent loss mechanism from saturating TLSs \cite{gao2008b,pappasGao2011,wangMartinis2009,sageWelander2011} described by:
\begin{equation}
Q_\text{TLS}(\bar{n},T) = \frac{Q_{\text{TLS},0}}{\tanh{\frac{\hbar\omega}{kT}}}\sqrt{1 + \left(\frac{\bar{n}}{n_c}\right)^\beta \tanh{\frac{\hbar\omega}{kT}}}
\end{equation}
Here $Q_{\text{TLS},0}$ is the inverse linear absorption from
TLSs, $\omega$ is the resonant frequency, and $\beta$ and $n_c$ are parameters characterizing TLS saturation \cite{gao2008b,sageWelander2011}.
Nonlinear effects (from kinetic inductance) \cite{anferovSchuster2020} limit the power range where linear measurements can be performed, however at high powers we observe that $Q_i$ begins to saturate, indicating the presence of other loss mechanisms.

By examining the power and temperature dependence of $Q_i$, we can further distinguish between sources of loss.
The full behavior is captured with a model that includes TLS loss ($Q_\text{TLS}$) \cite{gao2008b,pappasGao2011,wangMartinis2009,sageWelander2011}, equilibrium quasiparticle loss ($Q_\sigma$) \cite{reagor2016,tinkham1996,mattisBardeen1958} and other loss mechanisms that are  power and temperature-independent ($Q_\text{other}$):
\begin{equation}
\frac{1}{Q_i(T,\bar{n})} = \frac{1}{Q_\text{TLS}(\bar{n},T)} + 
\frac{1}{Q_\sigma(T)} + \frac{1}{Q_\text{other}}
\label{eqn_losses}
\end{equation}
The quasiparticle loss term is parameterized by:
\begin{equation}
Q_\sigma(T) = Q_{\sigma_0}\frac{\sigma_2(T,T_c)}{\sigma_1(T,Tc)}
\end{equation}
where $\sigma_1$ and $\sigma_2$ respectively are the real and imaginary parts of the complex surface conductance, calculated by numerically integrating the Mattis-Bardeen equations for $\sigma_1/\sigma_n$ and $\sigma_2/\sigma_n$, and $\sigma_n$ is the normal conductance. \cite{reagor2016,tinkham1996,mattisBardeen1958}.

To investigate the effects of quasiparticle loss, we measure $Q_i$ with TLS loss mostly saturated (at $\bar{n}\sim10^5$) for several representative resonators as a function of temperature. 
We show the results in \Fref{fig9}c along with $Q_\sigma(T)$ and fits to the full model described above.
Below approximately $1.5~$K we find that $Q_i$ is almost unaffected by $Q_\sigma$ and is nearly temperature independent for most devices.
As the temperature approaches a significant fraction of niobium's critical temperature ($9.2~$K), $Q_\sigma$ rapidly becomes dominant.
However, at the low temperatures relevant for quantum experiments, $Q_\sigma$ exceeds measured values of $Q_i$ by several orders of magnitude, suggesting that quasiparticle loss contributions are negligible in this regime.

Having determined that the dominant loss contributions come from power-independent loss $Q_\text{other}$ and $Q_\text{TLS}$, 
we can neglect thermal contributions to $Q_i$ at low temperatures.
Upon inspection of the power dependence of a typical resonator in \Fref{fig9}b, we find the increase of $Q_i$ from TLS saturation is relatively small, unlike what is seen in many microwave loss studies \cite{crowleydeLeon2023,verjauwRadu2021}.
Using the model above, we find $Q_{\text{TLS},0}=0.953\times10^6$ while $Q_{\text{other}}=1.17\times10^5$, indicating that TLSs are not the dominant source of loss.

For one resonator in particular we observe quality factors significantly higher than average, providing additional insight into loss origins.
This device falls very close in frequency to a nearby higher-linewidth resonance, which results in more asymmetric coupling as shown in \Fref{fig9}d, and likely modifies its radiation properties \cite{Ren2022quasinormal}.
We find Equation \ref{eq_s21} is able to accurately capture $Q_i$ from the response, but unlike the other resonators, the power dependence shown in \Fref{fig9}e is much more pronounced.
For this device, we find single-photon $Q_i=0.827\times10^6$, which is similar to state-of-the-art microwave resonators \cite{verjauwRadu2021, crowleydeLeon2023}, and the loss sources can be disentangled into $Q_{\text{TLS},0}=1.03\times10^6$ and $Q_{\text{other}}=4.18\times10^6$.
This independent loss limit is significantly higher than in most of our resonators, and $Q_i$ is instead primarily dominated by $Q_{\text{TLS}}$.
The latter is consistent with the other resonators on chip, giving us a better insight into millimeter-wave TLS loss.

\subsection{Reducing Losses With Surface Oxide Etch}

When exposed to air, niobium is known to slowly evolve a lossy amorphous surface oxide layer containing dissipative sub-oxides \cite{changYoon2015} and TLSs \cite{verjauwRadu2021}.
Surface treatments are commonly used to remove this surface layer in microwave resonators to reduce loss \cite{verjauwRadu2021}. Although TLS density has not yet been investigated in the W-band, losses from niobium sub-oxides are believed to be more pronounced at higher frequencies \cite{changYoon2015}.
To study the effect of surface processing on $Q_i$, we repeat the measurements summarized in Fig. \ref{fig9}a-b for devices that underwent different aging times and etch conditions.  In Fig. \ref{fig9}f,
we plot the low and high power limits of measured $Q_i$ as well as the fitted value of $Q_{\text{TLS},0}$ for devices from five separate chips. 

Between samples A and B we observe that 5 days of aging reduces both $Q_{\text{TLS},0}$ and $Q_\text{other}$, leading to lower quality factors for resonators exposed to air for several days, which is consistent with niobium oxide regrowth that has been shown to increase loss in microwave devices \cite{verjauwRadu2021}.
This can be mitigated by selectively removing the surface layer of niobium oxide after fabrication using a buffered solution of hydrogen fluoride (BHF) \cite{verjauwRadu2021}, which we achieve by immersing samples C, D and E in a 5\% BHF solution for $40~$min immediately prior to mounting and measurement.

\begin{figure}[t]
\centering
\includegraphics[width=2.5in]{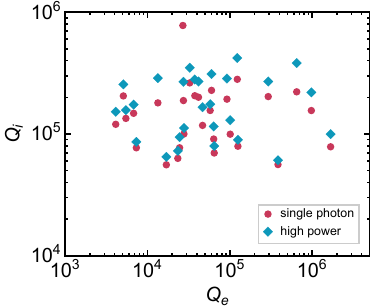}
\caption{Resonator loss (internal Q) for single-photon (red) and high-powers (blue) as a function of coupling Q, showing no correlation and confirming coupling does not increase loss.}
\label{fig10}
\end{figure}

We observe that this BHF treatment can reverse the effects of air exposure in Sample C, which despite experiencing a longer air exposure of 8 days, has higher average $Q_{\text{TLS},0}$ and $Q_{\text{other}}$ than Sample B after the BHF treatment.
Applying this surface treatment to samples D and E, which experienced reduced initial air exposure yield consistently lower losses than samples A-C, including the highest-Q resonator in this study described above.
Combining minimal air exposure with surface oxide removal using BHF in sample E, we are able to consistently obtain resonators with single photon internal quality factors above $1.4\times10^5$, and an average $Q_{\text{TLS},0}=1.04\times10^6$: significantly higher than previously measured for planar millimeter-wave devices \cite{stokowskiSafavi-Naeini2019,anferovSchuster2020,gaoLeduc2009,shirokoffZmuidzinas2012,endoKlapwijk2013,u-yenWollack2017}.

While these values of loss are much closer to those reported for microwave devices \cite{verjauwRadu2021}, in the power-dependent measurements above we observe that our millimeter-wave resonators are on average limited by power and temperature-independent loss $Q_\text{other}$ to a much greater extent than TLS loss, which limits microwave devices \cite{crowleydeLeon2023,verjauwRadu2021}.
This loss could come from a variety of sources, including remnants of conduction loss from the copper enclosure \cite{huangOliver2021}, seam loss from an imperfect seal between the halves of the enclosure \cite{brechtSchoelkopf2015}, radiation loss \cite{catelaniGlazman2011} or additional power-independent microscopic relaxation channels such as conductive loss in the niobium sub-oxides \cite{verjauwRadu2021}.
To estimate the impact of radiation losses, we can verify that our ground-shielded resonator design protects the resonance from radiative loss induced by coupling.
In \Fref{fig10} we plot $Q_i$ for single-photon and high-power limits ($Q_\text{other}$) as a function of $Q_e$, and observe no correlation with either.
Thus we can design a resonant circuit with a wide range of coupling strengths without affecting coupling to lossy or radiative channels.
As a final note, an examination of all the devices in \Fref{fig9}f, shows that on average $Q_{\text{TLS},0}$ and $Q_\text{other}$ scale similarly relative to each other when affected by aging and surface treatment.
This is highly suggestive that the remaining millimeter-wave source of loss is still tied to materials-induced decoherence in the superconductor surface, and warrants further studies.

\section{Conclusion}
We have demonstrated an on-chip millimeter-wave resonator design with a ten-fold improvement in loss over previous work, and leveraged this platform to investigate sources of single-photon decoherence in the W band, finding that scaling up superconducting device frequency does not significantly change the nature of decoherence.
Using a specifically-designed waveguide to slotline transition based on a finline taper, we demonstrate a packaging technique for probing on-chip devices at high frequencies which improves on dipole coupling techniques previously used to address millimeter-wave resonators.
The increased control afforded by this coupling method enables more complex device designs with high-frequency signal routing, and we show that our coupling strengths can be adjusted over a wide range without introducing circuit losses.

Having shown that planar millimeter-wave resonators compatible with planar fabrication techniques can achieve performance comparable to microwave quantum circuits \cite{crowleydeLeon2023,verjauwRadu2021}, we now have suitable infrastructure to develop more advanced superconducting devices for astronomical detection and studying complex spin systems in this frequency regime.
This platform could be further enhanced by incorporating materials with improved surface oxide characteristics \cite{crowleydeLeon2023}, and the operating range could be extended to even greater temperatures with even higher-$T_c$ materials such as niobium nitride or niobium titanium nitride.
Furthermore, introducing a high-frequency nonlinear element \cite{anferovSchuster2023b,anferovSchuster2020,kimSemba2021} into our design would enable a millimeter-wave artificial atom operating at liquid-helium-4 temperatures, paving the way for a new generation of high-frequency higher-temperature quantum tools.

\ack
The authors thank P. Duda for assistance with fabrication process development, and M. W., A. Kumar and A. Suleymanzade for useful discussions.
This work is supported by the U.S. Department of Energy Office of Science National Quantum Information Science Research Centers as part of the Q-NEXT center, and partially supported by the University of Chicago Materials Research Science and Engineering Center, which is funded by the National Science Foundation under Grant No. DMR-1420709.
This work made use of the Pritzker Nanofabrication Facility of the Institute for Molecular Engineering at the University of Chicago, which receives support from Soft and Hybrid Nanotechnology Experimental (SHyNE) Resource (NSF ECCS-2025633).

\bibliographystyle{iopart-num}
\bibliography{ms}

\appendix
\section{Device Fabrication}
\qty{100}{\um}-thick C-plane polished sapphire wafers are annealed at $1200^\circ~$C for 1.5 hours, and allowed to slowly cool to room temperature.
Next the wafers are ultrasonically cleaned in toluene, acetone, methanol, isopropanol and de-ionized (DI) water, and finally etched in a piranha solution kept at \qty{40}{\degreeCelsius} for 2 minutes and rinsed with de-ionized water.
Immediately following, the wafers are loaded into a Plassys MEB550S electron-beam evaporation system, where they are baked at $>$\qty{200}{\degreeCelsius} under vacuum for an hour to help remove water and volatiles.
When a sufficiently low pressure is reached ($<5\times 10^{-8}~$mBar), titanium is electron-beam evaporated to bring the load lock pressure down even further.
\qty{100}{\nm} of Nb is now deposited by first evaporating at $>0.5~$nm/s while rotating the substrate.
The substrate is allowed to cool in vacuum for several minutes, before exposing it to air.

The wafers are mounted on a silicon handle wafer using AZ1518 photoresist cured at \qty{115}{\degreeCelsius}, then coated with \qty{1}{\um} of AZ MiR 703 photoresist and exposed with a \qty{375}{\nm} laser in a Heidelberg MLA150 direct-write system.
The assembly is hardened for etch resistance by a \qty{1}{\min} bake at \qty{115}{\degreeCelsius} then developed with AZ MIF 300, followed by a rinse in DI water.
The resonators and/or finline structure is now etched in a chlorine inductively coupled plasma reactive ion etcher with Cl$_2$, BCl$_3$ and Ar.
The plasma conditions are optimized to be in the ballistic ion regime, which gives high etch rates with minimal re-deposition.
Immediately after exposure to air, the wafer is quenched in DI water: this helps prevent excess etching by quickly diluting any surface HCl (formed by adsorbed Cl reacting with water vapor in the air).
The remaining photoresist is thoroughly dissolved in a mixture of \qty{80}{\degreeCelsius} n-methyl-2-pyrrolidone with a small addition of surfactants, which also removes the substrate from the handle wafer.

With the superconducting structure complete, the wafer is ultrasonically cleaned with acetone and isopropanol, coated with a thick protective covering of photoresist (MiR 703) cured at \qty{115}{\degreeCelsius}, and diced into chips (see Table 1 for dimensions).
The protective covering is now dissolved in \qty{80}{\degreeCelsius} n-methyl-2-pyrrolidone with surfactants (we find this can also help remove stubborn organic residues from previous steps), and the chips are given a final ultrasonic clean with with acetone and isopropanol.
The finished chips are stored in air for variable time (see main text for discussion) then packaged and cooled down.

\section{Cryogenic Measurements and Calibration}
\subsection{Experimental Measurement Setup}
\begin{figure*}[ht!]
\centering
\includegraphics[width=5.1in]{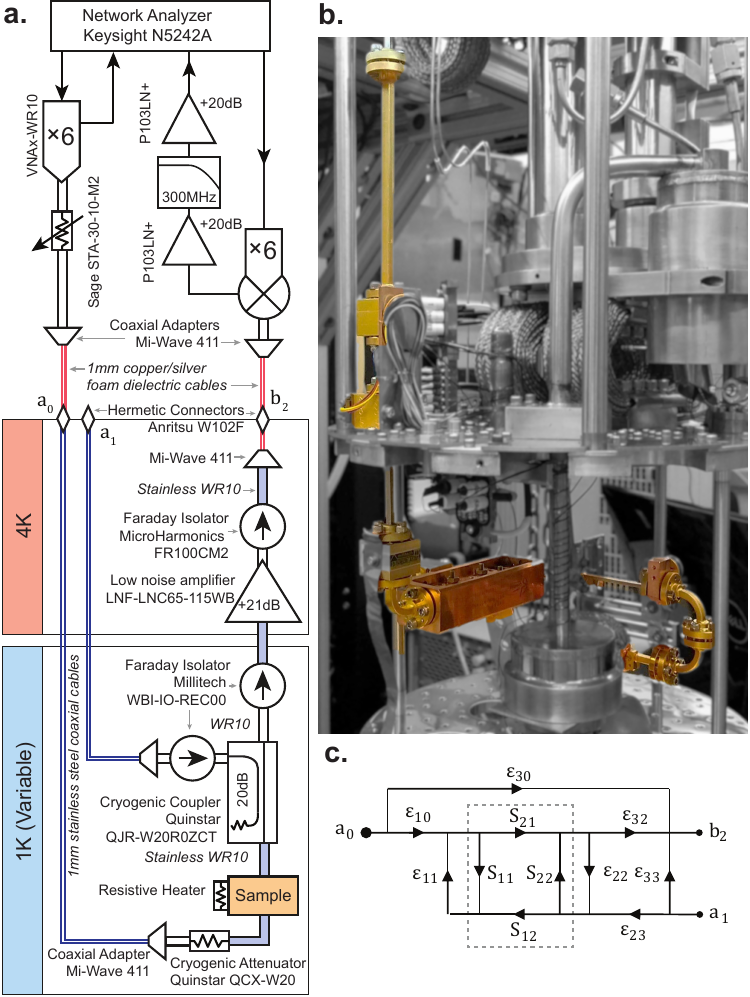}
\caption{\textbf{a)} Schematic of cryogenic millimeter-wave measurement setup. Colored tabs show temperature stages inside the Helium-4 adsorption refrigerator, which reaches a base temperature of $0.86~$K.
\textbf{b)} A photograph highlights relevant hardware inside the fridge.
\textbf{c)} Adjusted error network used for cryogenic TRL calibrations, where $S_x$ are the respective S-parameters of the sample, and the measurement paths $a_0$, $a_1$ and $b_2$ are also labelled in the schematic.
} 
\label{figS1}
\end{figure*}
All millimeter-wave characterization was performed in a custom built $^4$He adsorption refrigerator, with a base temperature of $0.86~$K, and a cycle duration of 3 hours. 
We generate millimeter-wave signals ($75$-$115~$GHz) at room temperature by sending microwave signals ($12$-$19~$GHz) into a frequency multiplier. 
The upconverted signal is sampled to establish a phase reference measurement.
We convert the generated waveguide TE$_{10}$ mode to a $1~$mm diameter stainless steel and beryllium copper coaxial cable, which carries the signal to the $1~$K stage of the fridge, thermalizing mechanically at each intermediate stage, then convert back to a WR-10 waveguide which leads to the device under test. 
The cables and waveguide-cable converters have a combined frequency-dependent loss ranging from $38.6~$dB to $49.8~$dB in the W-Band, dominated by the cable loss. 
In the case of a transmission measurement, the signal is further thermalized to $1~$K by a cryogenic 20 dB attenuator, and in the case of a reflection measurement, this thermalization is accomplished with a 20 dB cryogenic directional coupler with a copper body.
The sample is thermally isolated from the $1~$K stage of the refrigerator to allow local heating for temperature sweeps.

Wideband cryogenic millimeter-wave circulators are currently not commercially available, so instead a cryogenic directional coupler allows enables reflection measurements by allowing nearly all of the reflected and transmitted signal to pass through to a low noise amplifier.
Cryogenic faraday isolators minimize retro-reflections and prevent thermal radiation from leaking in on the output side, while still allowing good transmission.
Having passed outside the cryostat through custom-built hermetic adapters, the signal is downconverted and amplified for measurement.
The entire setup is summarized in \Fref{figS1}a-b.

\subsection{Cryogenic Calibration}
The cryogenic measurements described above introduce a complex network between the sample and the measurement equipment, a calibration must be performed to obtain accurate estimates of network parameters.
The necessity of attenuating components as well as active components with gain in the measurement chain make cryogenic calibrations a complex problem \cite{Yeh2013cryocal,Wang2021cryocal,Cataldo2015cryocal}.
Without access to a cryogenic millimeter-wave switch that can instantly select a calibration standard in-situ, we instead rely on a standard TRL-type calibration \cite{pozar2011} with carefully controlled sequential cooldowns with each standard.
The cryogenic measurement network can summarized with a partial network of error adapters in \Fref{figS1}c.

The input paths $a_0$ and $a_1$ both have significant attenuation ($\epsilon_{10}$ and $\epsilon_{23}$) to prevent room temperature noise from reaching the sample.
Because of this, signals directly reflecting off the samples (which are doubly attenuated) are too faint to measure, especially when overlaid on the imperfect return loss on some of the millimeter-wave components in the chain.
Instead, we use two separate input paths in order to characterize the response of the sample: when combined, this yields an error adapter network similar to the familiar two-port TRL network \cite{pozar2011}.

Aside from attenuating terms ($\epsilon_{10}$ and $\epsilon_{23}$) and the output amplification ($\epsilon_{32}$) which suffice for a simple correction, we must take into account several non-ideal terms in our error network.
As a consequence of highly attenuating inputs is that when normalized, the directivity terms ($\epsilon_{30}$ and $\epsilon_{33}$) are significantly more pronounced.
Additionally, the cryogenic attenuators and absorbers in our system appear have non-ideal return loss characteristics, resulting in non-negligible source match $\epsilon_{11}$ and even more pronounced directivity $\epsilon_{33}$.
Since we measure good return loss in the cryogenic faraday isolators used on the output line (see \Fref{figS1}a), the load match term $\epsilon_{22}$ should be significantly less pronounced.

As a result, we can simplify the procedure by neglecting $\epsilon_{22}$, and make a further assumption that our sample structure is symmetric (in practice this can be checked by reversing the sample direction).
These simplifications leave some unresolved frequency ripples on the measured spectrum (visible in Figure 5).
The measured transmission and reflection response is now simply:
\begin{equation}
    S_{21}^\text{M} = \frac{b_2}{a_0} = \epsilon_{30} + \frac{\epsilon_{10} \epsilon_{32} S_{21}}{1 - \epsilon_{11} S_{22}}
\end{equation}
\begin{equation}
    S_{22}^\text{M} = \frac{b_2}{a_1} = \epsilon_{33} + \left(S_{22} + \frac{\epsilon_{11}S_{21}^2}{1 - \epsilon_{11} S_{22}} \right) \epsilon_{23} \epsilon_{32}
\end{equation}
By performing measurements of through, reflect, and line standards which have known S-parameters \cite{pozar2011} we can solve this system of equations for relationship the error terms.
Combining this with direct measurements of each input line yields true values for each error term.
With the system now characterized, we can extract sample S-parameters from the transmission and reflection measurements above, giving a much more accurate picture.
Since each measurement now relies on a number of calibration experiments, this imparts a linear error (with typical vector magnitude between $0.03-0.09$) on the final complex S-parameter.
As the amount of measurement uncertainty now depends on the measurements themselves (defined by the relationships in Equation B.1 - B.2), this results in S-parameter uncertainty that increases as the S-parameter decreases, apparent in the significantly higher uncertainty on our reflection measurements.

\section{Leakage Bypassing the Transition}

\begin{figure}
\centering
\includegraphics[width=3.28in]{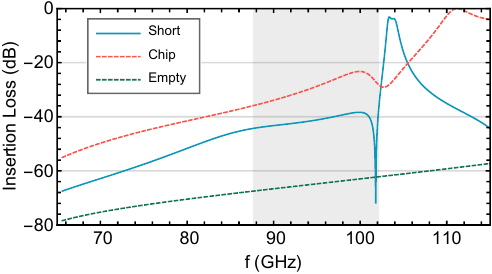}
\caption{Simulated transmission parameters for: a single transition terminated by an on-chip short, a sapphire chip with no metallization, and the structure with the chip removed. The operating band is highlighted in gray.}
\label{figS2}
\end{figure}

We have thus far treated the system of two back-to-back taper transitions as a single unified network described by a set of S-parameters.
However to verify the effectiveness of the transition with even greater precision, the transitions themselves could be de-embedded by defining an error sub-network inside the grey box in \Fref{figS1}c.
This could help with getting more accurate measurements of more complex on-chip devices, but in particular the isolation term of this sub-network (equivalent to $\epsilon_{30}$ in \Fref{figS1}c) could shed some light on how much signal passes through the slotline and reaches the resonators as opposed to bypassing the chip entirely.
With our measurement precision, the correction method described earlier already gives fairly high uncertainty without introducing these extra error parameters, so we are unable to directly measure this leakage with good accuracy.

However, we can use simulations (which show fairly good agreement with the measured responses) to estimate the relative magnitude of the leakage.
In \Fref{figS2}, the solid blue line shows the simulated transmission of a chip with a tapered transition on one side, and entirely covered by uniform ground plane on the other.
This measurement corresponds to the \textit{Reflect} standard used in the TRL calibration \cite{pozar2011}, and directly measures the isolation error term.
Thus we find that in the band of interest for our system (highlighted in \Fref{figS2}), the simulated leakage is below approximately 38~dB: this value results in a total reduction between 0.05 and 0.1~dB on our measured insertion loss: significantly lower than the uncertainty from calibration discussed in the previous section.

This leakage term is small enough to ignore for resonator measurements, however sheds light on imperfections in this design.
To estimate the leakage origin, we also simulate transmission of the geometry with a bare sapphire chip containing no metal, as well as transmission without the chip entirely, shown with dashed lines in \Fref{figS2}.
From this we conclude that the bare sapphire chip supports modes which help transmit significantly more signal than just the copper enclosure geometry itself.
While these spurious modes are suppressed by the surface metal (demonstrated by reduced transmission of the short) a more careful examination could help further improve the transition design.

\end{document}